\documentclass[seceq]{ptptex}

\title{
Nonextensive Thermodynamics of a Cluster consisting\\
of $M$ Hubbard Dimers ($M=1,2,3$ and $\infty$)
\footnote{Proceedings of CN-Kyoto, March 14-18, 2005}
}

\author{
Hideo Hasegawa
\footnote{e-mail:  hasegawa@u-gakugei.ac.jp}
}

\inst{
Department of Physics, Tokyo Gakugei University, \\
Koganei, Tokyo 184-8501, Japan
}

\recdate{Mmmmm DD, 2005}

\abst{
The thermodynamical property of a small cluster
including $M$ Hubbard dimers, 
each of which is described
by the two-site Hubbard model, has been discussed 
within the nonextensive statistics (NES).
We have calculated
temperature and magnetic-field
dependences of the specific heat and
susceptibility for $M = 1, 2, 3$ and $\infty$,
assuming the relation 
between $M$ and the entropic index $q$
given by $q=1+1/M$,
which was previously derived by several methods.
For relating the physical temperature $T$ to the Lagrange
multiplier $\beta$, two methods have been adopted:
$T=1/k_B \beta$ in the method A 
[Tsallis {\it et al.} Physica A {\bf 261} (1998) 534], 
and  $T=c_q/k_B \beta$ in the method B
[Abe {\it et al.} Phys. Lett. A {\bf 281} (2001) 126],
where $k_B$ denotes the Boltzman
constant, $c_q= \sum_i p_i^q$, and $p_i$ the 
probability distribution of the $i$th state.
A comparison between
the results calculated by the two methods
suggests that the method B may be more appropriate than the
method A for small-scale systems.
}

\begin{document}

\maketitle

\section{Introduction}


In the last several years,
much study has been made 
with the use of nonextensive statistics (NES)
which was initiated by Tsallis. 
\cite{Tsallis88}\tocite{NES}
Before discussing the NES,
let's recall the basic feature of the 
Boltzman-Gibbs statistics (BGS) 
for a system with internal energy $E$ and
entropy $S$, which is immersed in a large reservoir
with energy $E_0$ and entropy $S_0$.
The temperature of the small system $T$ is the same as
that of the reservoir $T_0$ where
$T=\delta E/\delta S$ and $T_0 = \delta E_0/\delta S_0$.
If we consider the number of possible microscopic states
of $\Omega (E_0)$ in the reservoir, its entropy
is given by $S_0 = k_B \:{\rm ln} \Omega(E_0)$ where
$k_B$ denotes the Boltzman constant.
The probability of finding the small system with the 
energy $E$ is given by 
$p(E)=\Omega(E_0-E)/\Omega(E) \sim {\rm exp}(-E/k_B T)$
with $E \ll E_0$.
When the physical quantity $Q$ 
of a system containing $N$ particles
is expressed by $Q\: \propto N^{\gamma}$,
it is classified into two groups in the BGS:
intensive ($\gamma=0$) or extensive one ($\gamma=1$).
The temperature and energy are typical
intensive and extensive quantities, respectively.
This is not the case in the NES, as will be shown below.

In the NES, on the contrary, 
the temperature of a nanosystem which is
in contact with the reservoir, is
expected to fluctuate around the temperature of the reservoir $T_0$
because of the smallness of nanosystems 
and their quasi-thermodynamical
equilibrium states with the reservoir.
Then the BGS distribution mentioned above has to be averaged
over the fluctuating temperature.
This idea has been expressed by 
\cite{Wilk00}\cite{Beck02}\cite{Raja04}
\begin{eqnarray}
p(E) &=& \int_0^{\infty} \: d\beta \:e^{-\beta E}\: f^B(\beta) 
\nonumber \\
&=& [1-(1-q) \beta_0 E]^{\frac{1}{1-q}}
\equiv {\rm exp}_q (-\beta_0 \:E),
\end{eqnarray}
with 
\begin{eqnarray}
q&=& 1 + \frac{2}{N}, \\
f^B(\beta)&=& \frac{1}{\Gamma \left( \frac{N}{2} \right)}
\left( \frac{N}{2\beta_0} \right)^{\frac{N}{2}}
\beta^{\frac{N}{2}-1} 
{\rm exp}\left( -\frac{N \beta}{2\beta_0} \right), \\
\beta_0 &=& \frac{1}{k_B T_0}
= \int_0^{\infty} \:d\beta  f(\beta)\:\beta \equiv E(\beta), \\
\frac{2}{N}&=& \frac{E(\beta^2)-E(\beta)^2}{E(\beta)^2},
\end{eqnarray}
where ${\rm exp}_q(x)$ denotes the $q$-exponential function
defined by
\begin{eqnarray}
{\rm exp}_q(x)&=&[1+(1-q)x]^{\frac{1}{1-q}},
\hspace{1.0cm}\mbox{for $1+(1-q)x > 0$} \nonumber \\
&=& 0, 
\hspace{4.0cm}\mbox{otherwise} 
\end{eqnarray}
$q$ expresses the entropic index, 
$E(Q)$ the expectation value of $Q$
averaged over the $\Gamma$ (or $\chi^2$) distribution 
function $f^B(\beta)$, 
$\beta_0$ the average of the fluctuating $\beta$
and $2/N$ its dispersion.
The $\Gamma$ distribution is emerging from the sum
of squares of $N$ Gaussian random variables,
related discussion being given in Sec. 3.

The functional form of the probability distribution
$p(E)$ expressed by Eq. (1.1) was originally
derived by the maximum 
entropy method \cite{Tsallis88}\cite{Tsallis98}
with the generalized entropy given by
\begin{equation}
S_q= k_B \left( \frac{\sum_i p_i^q-1}{1-q} \right)
= - k_B \sum_i \:p_i^q \:{\rm ln}_q(p_i),
\end{equation}
where 
$p_i$ [$=p(\epsilon_i)$] denotes the probability 
distribution for the energy $\epsilon_i$ in the system,
and ${\rm ln}_q(x)$ [$=(x^{1-q}-1)/(1-q)$] the $q$-logarithmic 
function, the inverse of the $q$-exponential function.
The important consequence of the NES is
that entropy and energy 
are not proportional to $N$ in nanosystems.

In our previous papers \cite{Hasegawa04,Hasegawa05} 
(referred to as I and II, respectively),
we have applied the NES to
the Hubbard model, which is
one of the most important models in solid-state physics.
Thermodynamical properties of canonical \cite{Hasegawa04,Hasegawa05} 
and grand-canonical ensembles \cite{Hasegawa05}
of a Hubbard dimer described by the two-site Hubbard
model have been calculated within the NES.
It has been shown that the specific heat and susceptibility
calculated by the NES 
may be significantly different from those calculated
by the BGS when the entropic index $q$ 
departs from unity, the NES with $q=1$ reducing to
the BGS. 

We will consider in the present paper,
a nanocluster containing multiple ($M$) Hubbard dimers 
in order to discuss the $M$ dependence 
of their thermodynamical properties. We have assumed
the $M-q$ relation:
\begin{equation}
q =1+\frac{1}{M},
\end{equation}
which is derived from Eq. (1.2) with $N=2 M$.

The paper is organized as follows. After discussing the
adopted model and calculation method, 
we will present in \S 2,
numerical calculations of temperature and
magnetic-field dependences of thermodynamical quantities
for various $M$ values.
The final \S 3 is devoted to discussion and conclusion.

\section{Nonextensive statistics for Hubbard dimers}

\subsection{Adopted model and calculation method}

We have adopted canonical ensembles 
of a small cluster containing $M$ Hubbard dimers,
each of which is described by the two-site Hubbard model.
Interdimer interactions are assumed to be negligibly small.
The Hamiltonian is given by
\begin{eqnarray}
H &=& \sum_{\ell=1}^M \: H_{\ell}^{(d)}, \\
H_{\ell}^{(d)} &=& -t \sum_{\sigma}
( a_{1\sigma}^{\dagger} a_{2\sigma} 
+  a_{2\sigma}^{\dagger} a_{1\sigma}) 
+ U \sum_{j=1}^2 n_{j \uparrow} n_{j \downarrow }
- \mu_B B \sum_{j=1}^2 (n_{j\uparrow} - n_{j \downarrow}), \nonumber\\
&& \hspace{8cm} \mbox{($1, 2 \in \ell$)}
\end{eqnarray}
where $H_{\ell}^{(d)}$ denotes the Hamiltonian for the $\ell$th dimer,
$n_{j\sigma}
= a_{j\sigma}^{\dagger} a_{j\sigma}$,
$a_{j\sigma}$ an annihilation operator of an electron with
spin $\sigma$ on a site $j$ ($\in \ell$), 
$t$ the hopping integral,
$U$ the intraatomic interaction,
$\mu_B$ the Bohr magneton and 
$B$ an applied magnetic field.
Six eigenvalues of $H_{\ell}^{(d)}$ are given by
\begin{equation}
\epsilon_i=0, \; 2 \mu_B B, \; -2 \mu_B B, 
\; U, \; \frac{U}{2}+\Delta, \; \frac{U}{2}-\Delta,
\hspace{1cm}\mbox{for $i=1-6$}
\end{equation}
where $\Delta=\sqrt{U^2/4+4 t^2}$.
\cite{Bernstein74}\cite{Suezaki72}
The number of eigenstates of the total Hamiltonian 
$H$ is $6^M$.

The entropy $S_q$ in the Tsallis NES
is defined by \cite{Tsallis88}\cite{Tsallis98}
\begin{equation}
S_q=k_B \left( \frac{{\rm Tr} \:(\rho_q^q) - 1}{1-q} \right).
\end{equation}
Here $\rho_q$ stands for 
the generalized canonical density matrix,
whose explicit form will be determined shortly [Eq. (2.7)].
We impose the two constraints given by 
\begin{eqnarray}
{\rm Tr} \:(\rho_q)&=&1, \\
\frac{{\rm Tr} \:(\rho_q^q H)}{{\rm Tr} \:(\rho_q^q)}
&\equiv& <H>_q = E_q,
\end{eqnarray}
where the normalized formalism is adopted. \cite{Tsallis98}  
The variational condition for the entropy with
the two constraints given by Eqs. (2.5) and (2.6)
yields
\begin{equation}
\rho_q=\frac{1}{X_q} {\rm exp}_q 
\left[ -\left( \frac{\beta}{c_q} \right) (H-E_q) \right],
\end{equation}
with
\begin{equation}
X_q={\rm Tr}\: \left( {\rm exp}_q 
\left[-\left( \frac{\beta}{c_q} \right) (H-E_q)\right] \right),
\end{equation}
\begin{equation}
c_q= {\rm Tr} \:(\rho_q^q) = X_q^{1-q},
\end{equation}
where ${\rm exp}_q [x]$
expresses the $q$-exponential function defined by Eq. (1.6)
and $\beta$ is a Lagrange multiplier:
\begin{equation}
\beta=\frac{\partial S_q}{\partial E_q}.
\end{equation}
Specific heat and suscptibility 
have been calculated in I and II.

For relating the physical temperature $T$  
to the Lagrange multiplier $\beta$,
we have adopted the two methods A and B, given by \cite{Hasegawa05} 
\begin{eqnarray}
T &=& \frac{1}{k_B \beta},
\hspace{2cm}\mbox{(method A)}\\
&=& \frac{c_q}{k_B \beta}.
\hspace{2cm}\mbox{(method B)}
\end{eqnarray}
The method A proposed in Ref. 2 
is the same as the extensive BGS.
The method B is introduced so as to satisfy the {\it zero}th law
of thermodynamical principles and the generalized Legendre
transformations. \cite{Abe01}
It has been demonstrated that the negative specific heat 
of a classical gas model which is realized in the method A, \cite{Abe99}
is remedied in the method B. \cite{Abe01}
A difference between the two methods does not matter
as far as we consider only the non-exponential distribution
in the NES.  It yields, however, a significant difference
in the temperature dependence of thermodynamical
quantities, as will be shown in the following subsection.

\subsection{Numerical calculations}

\subsubsection{Temperature dependence}

In order to study how thermodynamical quantities of a cluster
with Hubbard dimers depend on its size $M$, we have made 
numerical calculations, assuming the $M-q$ relation given 
by Eq. (1.8), where results for $M=\infty$ correspond to 
those of the BGS ($q=1$).
Figures 1(a)-1(d) show the results for $U/t=5$. 
The specific heat and susceptibility
shown in Figs. 1(a) and 1(b), have been
calculated by the method A 
with $q=$ 2.0, 1.5, and 1.333 for
$M=1$, 2 and 3, respectively.
Figures 1(c) and 1(d) express 
$C_q$ and $\chi_q$, respectively, calculated by the method B.
We note that
physical quantities in a small cluster with $M \sim 1-3$ are 
rather different from
those of bulk-like systems with $M=\infty$,
although properties of clusters gradually approach
those of bulk with increasing $M$.

We note in Figs. 1(a)-1(d) that by varying $M$, the maximum
values of the specific heat ($C^*_q$) and 
the susceptibility ($\chi^*_q$) and corresponding
temperatures of $T^*_C$ and $T^*_{\chi}$ are changed.
Figure 2(a) shows $T^*_C$ and $T^*_{\chi}$, 
and Fig. 2(b) depicts $C^*_q$ and $\chi^*_q$,
which are plotted against $1/M$:
solid and dashed lines denote results 
calculated by the methods A and B, respectively.
It is shown in Fig. 2(a) that with increasing $1/M$,
$T^*_{\chi}$ calculated by the method A
is much increased than that calculated by the method B.
We note also that with increasing $1/M$,
$T^*_{C}$ of the method B is increased
while that of the method A is decreased.
Figure 2(b) shows that $C_q^*$ in the method A is smaller
than that in the method B, whereas
$\chi^*_q$ in the method A is the 
same as that in the method B.

\subsubsection{Magnetic-field dependence}

From the $B$ dependence of the six eigenvalues
of $\epsilon_i$ [Eq. (2.3)],
we note the crossing of the eigenvalues
of $\epsilon_3$ and $\epsilon_6$ at the critical filed:
\begin{equation}
\mu_B B_c = \sqrt{\frac{U^2}{16}+ t^2}-\frac{U}{4},
\end{equation}
leading to $\mu_B B_c/t=0.351$ for $U/t=5$.
For $B < B_c$ ($B > B_c$),
$\epsilon_6$ ($\epsilon_3$) is the ground state.
At $B \sim B_c$ the magnetization $m_q$ is rapidly increased
as shown in Figs. 3(a) and 3(b) for $k_B T/t=1.0$
and 0.1, respectively:
the transition at lower temperature is more
evident than at higher temperature. 
This level crossing also yields a peak in $\chi_q$ 
[Figs. 3(c) and 3(d)] and
a dip in $C_q$ [Figs. 3(e) and 3(f)].
It is interesting that
the peak of $\chi_q$ for $q=1.5$ in the NES 
is more significant than that
in the BGS whereas that of $C_q$ of the former
is broader than that of the latter. 
When the temperature becomes higher, these peak structures
become less evident.
Similar phenomenon in the field-dependent specific heat
and susceptibility have been pointed out in the
Heisenberg model within the BGS. \cite{Kuzmenko04}

Figure 3(a) and 3(b) remind us the quantum tunneling of magnetization
observed in magnetic molecular clusters such as Mn4, 
Mn12 and Fe8. \cite{Mn12}
It originates from the level crossing of 
magnetic molecules which are parallel and anti-parallel
to the easy axis when a magnetic field is applied.

\section{Discussions and conclusions}

The $N-q$ relation given by $q=1+2/N$ [Eq. (1.2)] 
has been derived from 
the average of the BGS partition function
over the $\Gamma$ distribution $f^B$ given by Eq. (1.3).
By using the large-deviation approximation,
Touchette \cite{Touchette02} has obtained the 
alternative distribution
function $f^T(\beta)$ given by
\begin{eqnarray}
f^T(\beta)&=& \frac{\beta_0}{\Gamma 
\left( \frac{N}{2} \right)}
\left( \frac{N \beta_0}{2} \right)^{\frac{N}{2}}
\beta^{-\frac{N}{2}-2} 
{\rm exp}\left( -\frac{N \beta_0}{2\beta} \right).
\end{eqnarray}
Solid and dashed curves
in Fig. 4 express the $f^B$- and $f^T$-distribution functions,
respectively, for various $N$ values.
For $N \rightarrow \infty$,
both reduce to the delta-function densities, and
for a large $N=100$, both distribution functions
lead to similar results.
For a small $N \;(< 10)$, however, 
there is a clear difference between the two distribution 
functions.
We note that a change of variable $\beta \rightarrow \beta^{-1}$
in $f^T$ yields the distribution function similar to $f^B$.  
It should be noted that $f^T$ cannot lead to the
$q$-exponential function which plays a crucial role
in the NES. For a large $\epsilon$,
$f^T$ leads to the stretched exponential form of
$p(\epsilon) \sim e^{c \sqrt{\epsilon}}$
while $f^B$ yields the power form of
$p(\epsilon) \sim \epsilon^{-\frac{1}{q-1}}$.
This issue of $f^B$ versus $f^T$ is related to the {\it superstatistics},
which is currently studied with much interest.
\cite{Beck04}

Numerical calculations presented in the preceding
section have shown that
although results calculated by
the two methods A and B are qualitatively similar,
there are some quantitative difference,
as previously obtained in I and II. 
When we calculate
the Curie constant $\Gamma_q$ of the susceptibility
defined by $\chi_q(T)=(\mu_B^2/k_B)[\Gamma_q(T)/T]$,
the method A leads to anomalously large Curie constant
compared to that of the method B \cite{Hasegawa05}.
This agrees with the results for free spins
\cite{Hasegawa05}\cite{Mar00}
and for spin dimers described 
by the Heisenberg model \cite{Hasegawa06}.
A comparison between Eqs. (1.1) and (2.7) yield
the average temperature $<T>$ given by
\begin{equation}
\frac{1}{k_B <T>} \simeq \beta_o = \frac{\beta}{c_q}
\end{equation}
which is consistent with the method B.
These results suggest that the method B is 
more appropriate than the method A.
This is consistent with recent theoretical analyses 
\cite{Suyari05}\cite{Wada05}
[for a relevant discussion, see also Ref. 21].

In summary,
within the framework of the NES,
thermodynamical properties 
have been discussed  of
a cluster including $M$ dimers, each of which is described by
the two-site Hubbard model.
We have demonstrated that the thermodynamical properties
of small-scale systems are rather different
from those of bulk systems.
Owing to recent progress in atomic engineering, 
it is possible
to synthesize molecules containing relatively small numbers of 
magnetic atoms with the use of various methods
(for reviews, see Refs. 21-23). 
Theoretical and experimental
studies on nanoclusters with changing $M$
could clarify a link between the behavior of the
low-dimensional infinite systems and nanoscale finite-size systems.
The unsettled issues on $T-\beta$ 
and the $N-q$ relations in the current NES
are expected to be resolved by future theoretical and
experimental studies on nanosystems, which
are expected to be one of ideal systems
for a study on the NES.

\section*{Acknowledgements}
This work is partly supported by
a Grant-in-Aid for Scientific Research from the Japanese 
Ministry of Education, Culture, Sports, Science and Technology.  



\newpage

\begin{figure}
\caption{
The temperature dependences of (a) specific heat $C_q$ 
and (b) susceptibility $\chi_q$ (per dimer) 
of Hubbard dimers for $U/t=5$
calculated by the method A, and
those of (c) specific heat $C_q$
and (d) susceptibility $\chi_q$
calculated by the method B, with
$M=1$ (bold solid curves),
$M=2$ (chain curves), $M=3$ (dashed curves)
and $M= \infty$ (solid curves).
}
\label{fig1}
\end{figure}

\begin{figure}
\caption{
(a) $1/M$ dependence of the temperatures of 
$T^*_C$ (circles) and $T^*_{\chi}$ (squares) where 
$C_q$ and $\chi_q$ have the maximum values, respectively.
(b) $1/M$ dependence of the maximum values of 
$C^*_q$ (circles) and $\chi^*_q$ (squares).
Solid and dashed lines denote the results calculated 
by the methods A and B, respectively:
$T^*_{\chi}$ calculated by the method A shown in (a)
is divided by a factor of five.
}
\label{fig2}
\end{figure}

\begin{figure}
\caption{
The magnetic-filed dependence of
(a) the magnetization $m_q$ 
for $k_B T/t=1.0$ and (b) $k_B T/t=0. 1$,
(c) the susceptibility
for $k_B T/t=1.0$ and (d) $k_B T/t=0. 1$,
(e) the specific heat $\chi_q$ 
for $k_B T/t=1.0$ and (f) $k_B T/t=0. 1$,
with $U/t=5$ for $M=2$
calculated by the method A (solid curves)
and B (dashed curves) and 
for $M=\infty$ (chain curves).  
}
\label{fig3}
\end{figure}

\begin{figure}
\caption{
The distributions of $f^B(\beta)$ (solid curves) and
$f^T(\beta)$ (dashed curves) as a function of $\beta$
(see text).
}
\label{fig4}
\end{figure}

\end{document}